\documentclass[12pt]{article}

\usepackage{color}
\usepackage{cite}
\usepackage{graphicx}
\usepackage{caption}
\usepackage{amssymb}
\usepackage{pdfpages}

\newenvironment{sciabstract}{%
\begin{quote} \bf}
{\end{quote}}

\title{Activity-driven polymer knotting for macromolecular topology engineering} 

\author
{Jia-Xiang Li,$^{1,2}$ Song Wu,$^1$ Li-Li Hao,$^4$ Qun-Li Lei,$^{1,2,3\ast}$ and Yu-Qiang Ma$^{1,2,3\ast}$\\
\\
\normalsize{$^{1}$ National Laboratory of Solid State Microstructures and School of Physics,}\\
\normalsize{Collaborative Innovation Center of Advanced Microstructures,}\\ 
\normalsize{Nanjing University, Nanjing 210093, People's Republic of China}\\
\normalsize{$^{2}$ Jiangsu Physical Science Research Center,}\\
\normalsize{Nanjing 210093, People's Republic of China}\\
\normalsize{$^{3}$ Hefei National Laboratory,}\\
\normalsize{Hefei 230088, People's Republic of China}\\
\normalsize{$^{4}$ Research Institute for Biomaterials,}\\
\normalsize{Tech Institute for Advanced Materials,}\\
\normalsize{College of Materials Science and Engineering,}\\
\normalsize{Nanjing Tech University, Nanjing 211816, People's Republic of China}\\
\\
\normalsize{$^\ast$To whom correspondence should be addressed;}\\
\normalsize{E-mail: lql@nju.edu.cn (Q. L.); myqiang@nju.edu.cn (Y. M.).}
}

\date{}

\begin{document} 

\baselineskip24pt

\maketitle 

\section*{Abstract}

\begin{sciabstract}
Macromolecules can gain special properties by adopting knotted conformations, but engineering knotted macromolecules is a challenging task. Here we surprisingly find that knots can be efficiently generated in active polymer systems. When one end of an actively reptative polymer is anchored, it undergoes continual self-knotting as a result of intermittent giant conformation fluctuations and the outward reptative motion. Once a knot is formed, it migrates to the anchoring point due to a non-equilibrium ratchet effect. Moreover, when the active polymer is grafted on a passive polymer, it can function as a self-propelling soft needle to either transfer its own knots or directly braid knots on the passive polymer. We further show that these active needles can create inter-molecular bridging knots between two passive polymers. Our finding highlights the non-equilibrium effects in modifying the dynamic pathways of polymer systems, which have potential applications in macromolecular topology engineering.
\end{sciabstract}

\section*{Teaser}
An anchored active polymer can spontaneously form knots, providing an efficient way to engineer the topology of macromolecules.

\section*{Introduction}
Knot topology play an important role on the biochemical functions of macromolecules~\cite{Tubiana2024,Meluzzi2010,Suma2017, Saitta1999,Sulkowska2008,Christian2016,Yan2024}. For example,  knots can provide topological protection to proteins against ATP-based enzymatic degradation~\cite{SanMartin2017,Sriramoju2018}, or endow molecules with high selectivity for ion binding and notable catalytic activity~\cite{Fielden2017,Forgan2011}. Knotted structures have been identified in approximately 1\% of native proteins deposited in the Protein Data Bank (PDB)~\cite{Jamroz2015,Sulkowska2012,Lim2015}. Advanced synthetic chemistry enables the creation of various complex knotted molecules~\cite{Danon2017,Leigh2020}. However, controlling the knot topology on linear biomacromolecules were proven to be challenging~\cite{Polles2015,King2010}, despite substantial efforts in macromolecular topology engineering~\cite{Arai1999,Qu2021, DabrowskiTumanski2017,Jackson2017}.

It is known that an infinitely long polymer in equilibrium inevitably encounters knotting due to thermal fluctuations~\cite{Summers1988}. The knotting probability is affected by factors such as local environments, chain sequence and flexibility~\cite{Micheletti2011,DAdamo2015,Coluzza2013,Lu2024}. However, the probability of spontaneous knotting  is particularly low, and the knotting is reversible~\cite{Koniaris1991,Tubiana2013}. Macroscopically, it has been demonstrated that a classical chain of beads can undergo spontaneous knotting under external driving such as shaking and agitating~\cite{Raymer2007, Belmonte2001,Hickford2006}, which suggests that non-equilibrium forces may be used to manipulate the topology of macromolecules microscopically~\cite{Favier2021,Amin2018}. Active polymers are chain structures driven out of equilibrium by local active forces~\cite{Winkler2017,Deblais2020,Juelicher2007,Kurzthaler2021,Feng2023}. At the molecular level, examples of active polymers include nucleic acids under the action of polymerases or helicases~\cite{Guthold1999,Stano2005}, actin filaments and microtubules pulled by myosin or kinesin~\cite{Yanagida1984,Howard1989}, flagella driven by proton pump~\cite{Glagolev1978}. Macroscopic active polymers include snakes, worms and artificial soft robots~\cite{Dreyfus2005,Biswas2017,Zhu2024}. These active polymers exhibit unusual dynamics behaviors inaccessible for their passive counterpart, like active reptation (railway motion)~\cite{IseleHolder2015,Li2023}, collapse without attraction~\cite{Bianco2018,Locatelli2021}, vortex or flapping~\cite{Chelakkot2014,Pochitaloff2022,Anand2019}. Nevertheless, how activity affects the knotting of macromolecules remains largely unexplored~\cite{Vatin2024,Vatin2024_2}.

In this study, we find that when one end of an actively reptative polymer chain is anchored, the polymer can undergo continual self-knotting: once a knot is formed, it migrates to the anchoring point and accumulates to form multiple knots configuration. The knotting rate is found to be controlled by both the activity strength and persistent length of the polymer. We find that the intermittent giant conformation fluctuations and the outward reptative motion of the active polymer are responsible for the self-knotting behaviors. These non-equilibrium effects also cause a  ratchet effect which leads to the knotting migration. Moreover, when grafted on the end of a passive polymer, the active polymer can function as a soft self-propelling needle to transfer its own knots to the passive polymer or directly braid knots on the passive polymer. We further show that the active needles can guide the inter-molecular bridge knotting between two oppositely anchored passive polymers. Our finding highlights the non-equilibrium effects in modifying the dynamic pathways of polymer systems, which have potential applications in macromolecular topology engineering and macromolecular braiding.

\section*{Results}

\begin{figure*} 
	\begin{center}
		\begin{tabular}{lr}
			\resizebox{160mm}{!}{\includegraphics{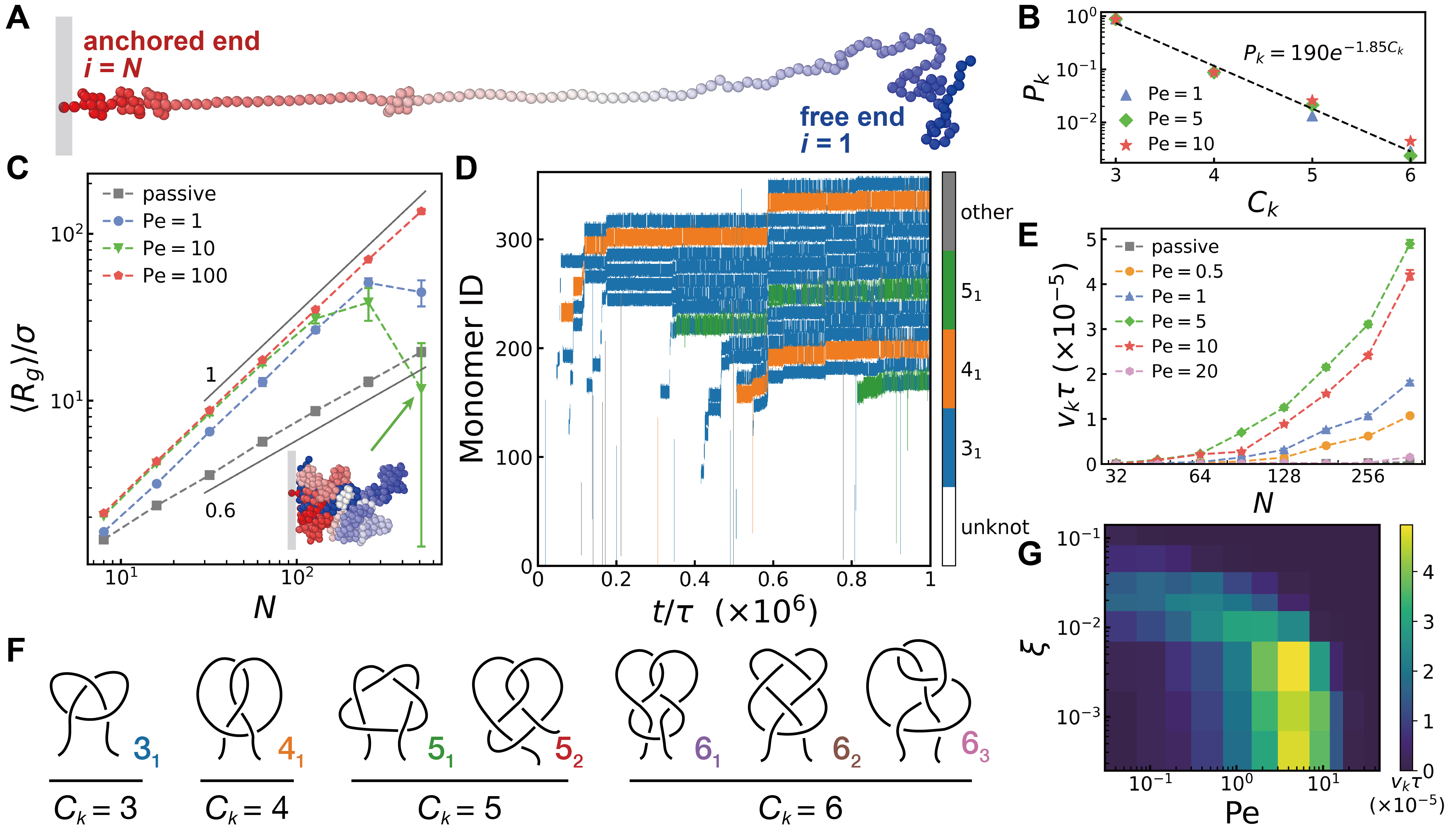}}
		\end{tabular}
	\end{center}
	\caption{\textbf{Self-knotting of an anchored active polymer.} (\textbf{A}) A snapshot of an anchored active polymer that forms four $3_1$ knots. (\textbf{B}) Knotting probability, $P_k$, as a function of the knot complexity represented by the minimum crossing number $C_k$, where the dashed line is a fitting of the data. (\textbf{C}) Averaged radius of gyration $\langle R_g\rangle$ as a function of chain length $N$, where the gray lines are guides for eyes with slopes of $0.6$ and $1$. The inset is a representative snapshot of a fully knotted and shrunken active polymer with $N=512$ and ${\rm Pe}=10$. (\textbf{D}) On a polymer chain with $N=362$ and ${\rm Pe}=1$, the evolution of the index of monomers involved in knots over time, where the color-bar represents knot type with representative projections shown in (\textbf{F}). (\textbf{E}) Knotting velocity $v_k$ as a function of chain length $N$ for polymers with different activities ${\rm Pe}$. (\textbf{G}) Phase diagram of active polymers ($N=362$) in ${\rm Pe}-\xi$ plane with different colors representing $v_k$.}
	\label{fg:model}
\end{figure*}

The system we consider is an actively reptative polymer chain consisting of $N$ monomers with one end (Monomer ID $i=N$) anchored on a surface (Fig.~\ref{fg:model}A). Each monomer sustains a constant self-propelling force with direction along the local tangent, pointing from monomer $i+1$ to monomer $i-1$. This kind of self-propelling is known to generate the active reptation or railway motion of the polymer in free 3D space~\cite{IseleHolder2015,Li2023}. Since the self-propelling force is outward, the  active reptation motion would stretch the anchored polymer chain. The mechanical stiffness and activity strength of the polymer are characterized by the relative rigidity $\xi$ and the P\'{e}clet number ${\rm Pe}$, whose definitions, as well as the model details, can be found in Methods.

For thermally equilibrated polymer chains, the Flory theory predicts a universal scaling law for polymer size (radius of gyration) $R_g\sim N^\nu$, where exponent $\nu\simeq0.6$ is applied for 3D self-avoiding free or surface-anchored chains~\cite{Flory1953,Degennes1980}. In Fig.~\ref{fg:model}C, we confirm the  $R_g\sim N^{0.6}$ scaling for the passive anchored chains. Nevertheless, for flexible active polymers ($\xi=0$) that experience tangentially outward propelling forces, when the activity is large (${\rm Pe}=100$),  exponent $\nu$ is found to approach 1,  indicating that the active chain is stretched straightly by the outward active forces. More interestingly, for systems with moderate activities ${\rm Pe}=1\sim 10$, we find a non-monotonic behaviors of $R_g(N)$, i.e, short polymers are straightened while long polymers are shrunken. In the inset of Fig.~\ref{fg:model}C, we show a snapshot of the equilibrated long active chain, which indicates the collapsing of the chain is due the formation of abundant knot structures. These knot structures were not found when the active reptative motion is inward~\cite{Wang2022}.

To study the knotting behaviors of the active polymer chain, we use a scanning strategy combined with Alexander polynomial to identify the topological state (knot type) of the polymer chain started from a straight initial configuration (see Fig.~S1 and Supplementary Note~3). In Fig.~\ref{fg:model}D, we depict the evolution of knot structures on a chain with ${\rm Pe=1}$, where each band represents the evolution of a knot on the chain, with color, location and width of the band denoting type, location and size of the knot, respectively. The representative projections of different knot types is shown in Fig.~\ref{fg:model}F. More details of the topology analysis can be found in Methods. As can be seen in Fig.~\ref{fg:model}D, multiple knots with different types can form on the chain. The complexity of the formed knots can be characterized by the minimum crossing number $C_k$~\cite{Adams1994}. In our simulations, we can observe all the seven types of knots with $C_k\leq6$ as shown in Fig.~\ref{fg:model}F, where the simplest $3_1$ knot is the most dominant knot type. For Gaussian chains, the probability of forming a specific knot, $P_k$, decreases exponentially with the increase of the knot complexity $C_k$~\cite{Shimamura2002}. This is because the formation of complex knots needs multiple twisting or loop-threading procedures. For biomacromolecules such as proteins, twisting is relatively easy, whereas threading is challenging~\cite{Taylor2007,Jackson2017}. Consequently, non-twist knots (i.e., $5_1$, $6_2$ and $6_3$) that require multiple threading procedures are rare. For our active polymer systems, we also find the exponential dependence of knotting rate on knot complexity (Fig.~\ref{fg:model}B). However, the easy observation of $5_1$, $6_2$, $6_3$ complex knots in our finite simulation time indicates that active reptation considerably enhances the threading efficiency of the polymer, making the weaving of complex knots possible.

During the  knotting process, once a knot is formed, it migrates to the anchored point in an intermittent manner. The number of knots on a polymer chain, $N_k$, increases first and then reaches a plateau after the length of knot-free tail being shorted than a critical value (Fig.~\ref{fg:model}D and  Fig.~S3).  We define the knotting rate $v_k$ as the slope of $\langle N_k\rangle(t)$ curves at the initial time (Fig.~S4), where the angle brackets denote ensemble average. In Fig.~\ref{fg:model}E, we plot the knotting rate $v_k$ as a function of chain length $N$ for polymers with different activities, from which one can see that the polymers with moderate activity (${\rm Pe}\approx5$) exhibit the highest knotting rate and the smallest critical chain length ($N\geq64$) to form knots.  In Fig.~\ref{fg:model}G, we summarize the knotting behaviors of the system in a phase diagram in dimensions of P\'{e}clet number $\rm Pe$ and relative rigidity $\xi$, which shows that flexible polymers ($\xi\leq0.02$) with moderate activities ($0.5\leq{\rm Pe}\leq20$) are most likely to exhibit self-knotting behaviors. Moderate rigidity ($0.01\leq\xi\leq0.05$) also facilitates the self-knotting of weak active chains (${\rm Pe}<0.5$), which is consistent with the knotting behanviors of passive polymers~\cite{Coronel2017,Zhu2021}.

\begin{figure*} 
	\begin{center}
		\begin{tabular}{lr}
			\resizebox{160mm}{!}{\includegraphics{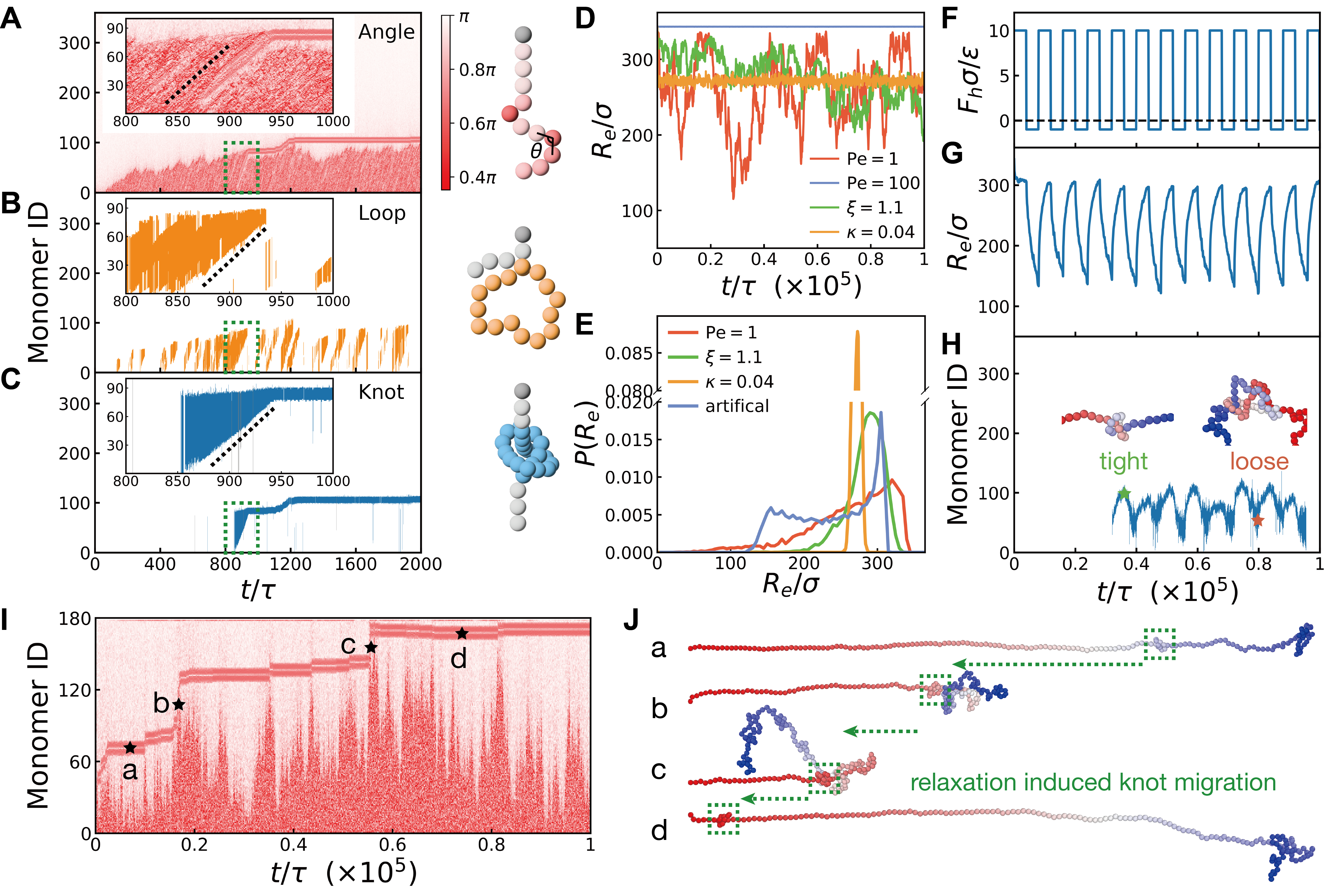}}
		\end{tabular}
	\end{center}
	\caption{\textbf{Mechanisms of knot formation and migration.} (\textbf{A}-\textbf{C}) Knot formation process of an active polymer with $N=362$ and ${\rm Pe}=5$ from a straightened initial configuration, where (\textbf{A}) displays the angle distribution, (\textbf{B}) shows the location of the loop structure and (\textbf{C}) demonstrates the location of the $3_1$ knot. The right panel of (\textbf{A}-\textbf{C}) shows the schematic diagrams of angle, loop and knot. (\textbf{D}) Knot migration process of a $3_1$ knot near the free tail on an active polymer with $N=362$ and ${\rm Pe}=1$. The figure shares the same color bar with (\textbf{A}). (\textbf{E}) The fluctuation and (\textbf{F}) the distribution of end-to-end distance $R_e$ for flexible active polymers with ${\rm Pe}=1$, ${\rm Pe}=100$, passive rigid polymers with $\xi=0.11$, and flexible polymers under external field $F_c=\kappa\varepsilon/\sigma$, where the chain length is $N=362$. (\textbf{G}) Representative snapshots labeled in (\textbf{D}) by black pentacles. (\textbf{H}) The strength of alternative external field exerted on the passive flexible polymers and (\textbf{I}) the resultant $R_e$ fluctuation, with corresponding $R_e$ distribution shown in (\textbf{F}). (\textbf{J}) Knotting behavior of the polymer under alternative external field, sharing the same color bar with Fig.~\ref{fg:model}D. The insets show representative snapshots of the knot structure labelled by pentacles.}
	\label{fg:loop}
\end{figure*}

We next study how an individual knot forms and migrates to the anchoring point of the active polymer. Since the formation of a knot structure is inevitably associated with local curvature of the polymer, we analyze the bond angle distribution on the chain. As shown in Fig.~\ref{fg:loop}A, starting from a straight initial conformation, curvature (corresponding to small bond angles) on the active polymer first occurs at the tip, subsequently spread towards the whole tail regime. The bond angle landscape exhibits tilted patterns with a tilt angle corresponding to the railway motion velocity of the free active polymer $v_0=F_a/\gamma$, where $\gamma$ is the frictional coefficient, as indicated by the black dashed line in the inset of Fig.~\ref{fg:loop}A. This suggests that the bending of polymer tail tip triggers the active reptation of the polymer leaded by the tip, resulting in the spread of curvature in the tail regime (see Movie~S1). In the recoiled tail, loop structures are frequently formed (Fig.~\ref{fg:loop}B). Here, a loop is defined as the chain regime between two non-adjacent monomers whose distance is less than $\sqrt{2}\sigma$ and monomer index distance is larger than $15$. The formed loops will be tightened and shrunk due to the railway motion of  the chain (inset of Fig.~\ref{fg:loop}B). Once the tail of the active polymer threads through the formed loop, a knot is formed and then tightened (Fig.~\ref{fg:loop}C). The typical threading dynamic process is shown in Movie~S2. This process resembles the process of tying a knot on macromolecules with optical tweezers~\cite{Arai1999}. However, it is not the general pathway of knotting in natural proteins, where the knot formation often involves a slipknot intermediate which can reduce the energy barrier required for threading~\cite{Noel2010,DabrowskiTumanski2018}. It is noteworthy that some generated knots have intrinsic chirality. For example, left-handed and right-handed $3_1$ knots are observed with statistically equal probability on the active polymers (see Fig.~S5). Knots with a preferred chirality may be generated by further introducing chiral motion into the active polymers. In this work, the knots are stretched so tightened by the active reptation that only 11 monomers are needed to form a $3_1$ knot, very close to the knot stretched by sufficiently large external forces~\cite{Caraglio2015}. Note that another knot tightening mechanism can exist in active circular polymer chains~\cite{Zhang2024}.

The analysis mentioned above reveals the dynamic pathway of knot formation. However, a fundamental question remains unanswered is why the active polymer chain prefers the self-knotting pathway, while its passive counterparts do not. From Fig.~\ref{fg:loop}A, one can see strong conformational fluctuations during the knotting process, which may be the clue to  this question. Therefore, in Fig.~\ref{fg:loop}E we plot the temporal evolution of the end-to-end distance $R_e$ of an active polymer (${\rm Pe}=1$) and compare it with its passive counterparts with a similar average $R_e$, including a passive semi-flexible polymer ($\xi=1.1$) and a flexible passive polymer under a uniform external field where each monomer experiences a constant force $F_c=\kappa\varepsilon/\sigma$. We find that the active polymer exhibits frequent transitions between shrunken and stretched states, generating giant fluctuations of $R_e$. Correspondingly, the distribution of $R_e$ of the active polymer is the broadest comparing with its passive counterparts (Fig.~\ref{fg:loop}F). For active polymers with large activity (${\rm Pe}=100$), conformational fluctuations are completely suppressed. $R_e$ appears as a perfectly straight line without self-knotting behaviors (Fig.~\ref{fg:loop}E), consistent with the results shown in Fig.~\ref{fg:model}C. To further confirm the role of intermittent giant conformation fluctuation in self-knotting, we artificially generate this fluctuation on passive polymers using alternative external field (See Fig.~\ref{fg:loop}H-I, Supplementary Note 1 and Movie~S3). This oscillating system resembles the driven hanging chain which also exhibits self-knotting behaviors~\cite{Belmonte2001}. We find that the self-knotting probability is indeed enhanced in this oscillating driven system (Fig.~\ref{fg:loop}J). However, the knot structures periodically loosen and tighten (Fig.~\ref{fg:loop}J), and finally become unfastened. In  contrast, the knots formed in active polymers are stretched tightened by the active reptation motion. All these results indicate that the intermittent giant conformation fluctuation and the active reptation of the active polymers are two ingredients for the irreversible self-knotting process.

To explore the migration of the knots on the active polymer, we perform simulations starting from an initially straight conformation with a $3_1$ knot near the tail. By drawing bond angle distribution in Fig.~\ref{fg:loop}D, we find that the knot moves on the polymer by intermittent hopping, which is accompanied by the curvature of the chain from the tail to the position of the knot (Movie~S4). The knot tightening and migrating mechanisms could be illustrated using Fig.~\ref{fg:loop}G. Each monomer in the active polymer experiences a propelling force along with the local tangent towards the free end. Therefore, when the active polymer is elongated, the knot on the polymer experiences an outward stretching force, which tightens the knot (Fig.~\ref{fg:loop}G(a)). In this stage, the knot exhibits no migration movement. When the free end of the active polymer retreats due to intermittent giant conformation fluctuations, the tension on the knot is reduced. Driven by local active forces, the polymer monomers continuously pass through the loosened knot. Consequently, the relative position of knot on the polymer chain shifts to the anchoring point (Fig.~\ref{fg:loop}G(b,c)). Once the active polymer is tightened again, the knot structure is fixed to the new position (Fig.~\ref{fg:loop}G(d)). A schematic is also drawn to illustrate this migration mechanism in Fig.~S6. Such sequential tightening-loosening-migrating-tightening events induce a ratchet effect that effectively prevents the unknotting and unidirectionally drives the knot migration. To prove this mechanism, we reverse the direction of propelling forces, i.e., $F_a=-F_a$. The migration direction is indeed reversed, and the migration velocity is substantially enhanced, close to the reptative speed of the active polymer, $v_0=F_a/\gamma$, due to the absence of tension and the jamming of the knot (see Fig.~S7 and Movie~S5). For comparison, the knots on passive polymers~\cite{Tubiana2013} or on mechanically stretched polymers~\cite{Narsimhan2016a,Renner2014,Soh2018} only randomly diffuse to one of the two ends. In all, the outward reptation and the giant conformation fluctuation impart a ratchet effect to the knots, resulting their unidirectionally migration along the backbone.

\begin{figure*} 
	\begin{center}
		\begin{tabular}{lr}
			\resizebox{160mm}{!}{\includegraphics{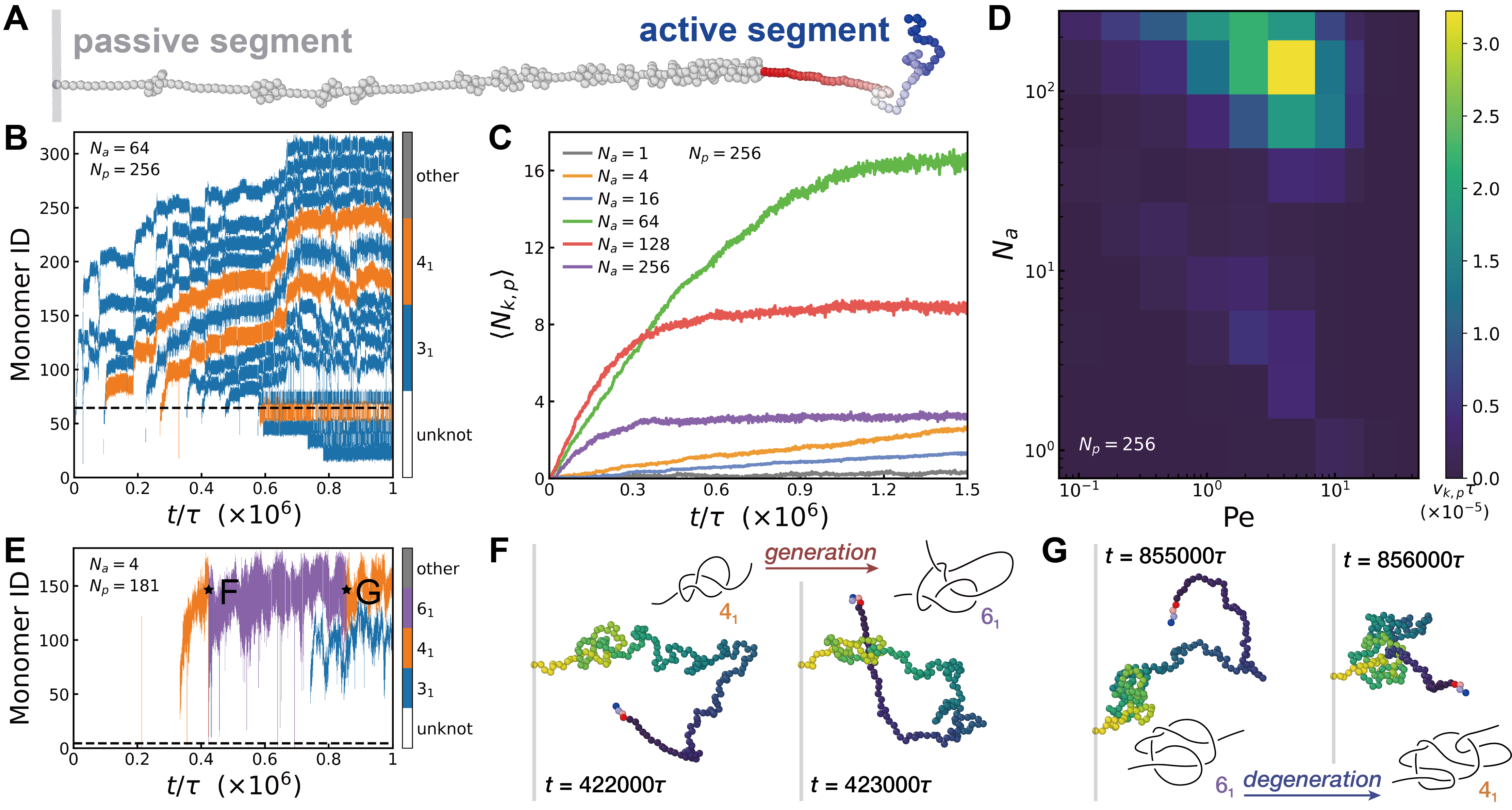}}
		\end{tabular}
	\end{center}
	\caption{\textbf{The active polymer as a needle for passive polymer knotting.} (\textbf{A}) A snapshot of a knotted active-passive copolymer and (\textbf{B}) the evolution of the knot structures on the copolymer with $(N_p, N_a)=(256, 64)$ and ${\rm Pe}=5$. (\textbf{C}) Average number of knots on the passive segment, $\langle N_{k,p}\rangle$, as a function of time. (\textbf{D}) Phase diagram of {knotting rate} on passive segment of the copolymer systems in ${\rm Pe}-N_a$ plane $v_k$. (\textbf{E}) The evolution of the knot structures on the copolymer with $(N_p, N_a)=(181, 4)$ and ${\rm Pe}=5$, where  the generation and degeneration of a $6_1$ knot are marked by two black pentacles and corresponding snapshot is shown in (\textbf{F}), (\textbf{G}) respectively. The black dashed lines in (\textbf{B}) and (\textbf{E}) show the boundary between the active segment and the passive segment.}
	\label{fg:needle}
\end{figure*}

We further consider an anchored passive polymer consists of $N_p$ monomers, whose tail is grafted with $N_a$ active monomers (Fig.~\ref{fg:needle}A). The reptative direction of the active polymer is outward. As shown in Fig.~\ref{fg:needle}B, we find that for this passive-active copolymer system with $(N_p,  N_a) =(256, 64)$ and ${\rm Pe}=5$, knots can form intermittently on active segment and be transferred to the passive segment rapidly. Different from the tight knots formed on long homogeneous active polymers (Fig.~\ref{fg:model}D), the knots on the passive segment can diffuse along the chain while repelling each other (Fig.~\ref{fg:needle}B), resembling the one-dimensional hard sphere gases~\cite{CardilloZallo2024}.

Similar to Fig.~\ref{fg:model}E, we extract the knotting rate on passive polymers, $v_{k,p}$, from the knot number on the passive segment as a function of time $\langle N_{k,p}\rangle(t)$, with representative data shown in Fig.~\ref{fg:needle}C. Then, a diagram of $v_{k,p}$ in dimensions of $\rm Pe$ and $N_a$ is plotted as shown in Fig.~\ref{fg:needle}D. We find that the knotting rate exhibits two hot spots in the phase space: a noticeable one at long active polymer length $N_a \simeq 100$ and a less-distinct one at short active polymer length $N_a \simeq 5$, both of which are at similar activity ${\rm Pe} \simeq 5$. This suggests that there are two scenarios of knotting on passive segment. When the length of the active segment is long, knots tend to initially form on the active segment and then migrate to the passive one. This scenario is characterized by rapid initial knotting rate similar to that observed in Fig.~\ref{fg:model}D. However, rapid knotting rate can lead to the formation of jammed multi-knot structures on the active segment which impedes knot migration to the passive segment. This results in an early saturation of knotting on the passive polymer for cases  $N_a \geq128$ in Fig.~\ref{fg:needle}C and Fig.~S8. On the other hand, when the length of the active segment is below a critical threshold ($N_p\lesssim20$), knots cannot form on the active segment alone. But the active segments can directly braid knots on the passive segment with a relatively lower knotting rate. The braided knots are loose (Fig.~\ref{fg:needle}E), which allows the further threading of the active segment, resulting in the generation and degeneration of complex knots in a short time window (e.g., $6_1$ knot in Fig.~\ref{fg:needle}F-G). The above finding indicates that active segments essentially act as self-propelling soft needles, which directs the knotting of passive polymer threads. Moreover, from Fig.~\ref{fg:needle}C, one can find that when $N_a=64$, the passive polymer can sustain the maximum number of knots. Thus, the most efficient knotting strategy is a balance between knotting rate and knot migrating rate.

\begin{figure*} 
	\begin{center}
		\begin{tabular}{lr}
			\resizebox{80mm}{!}{\includegraphics{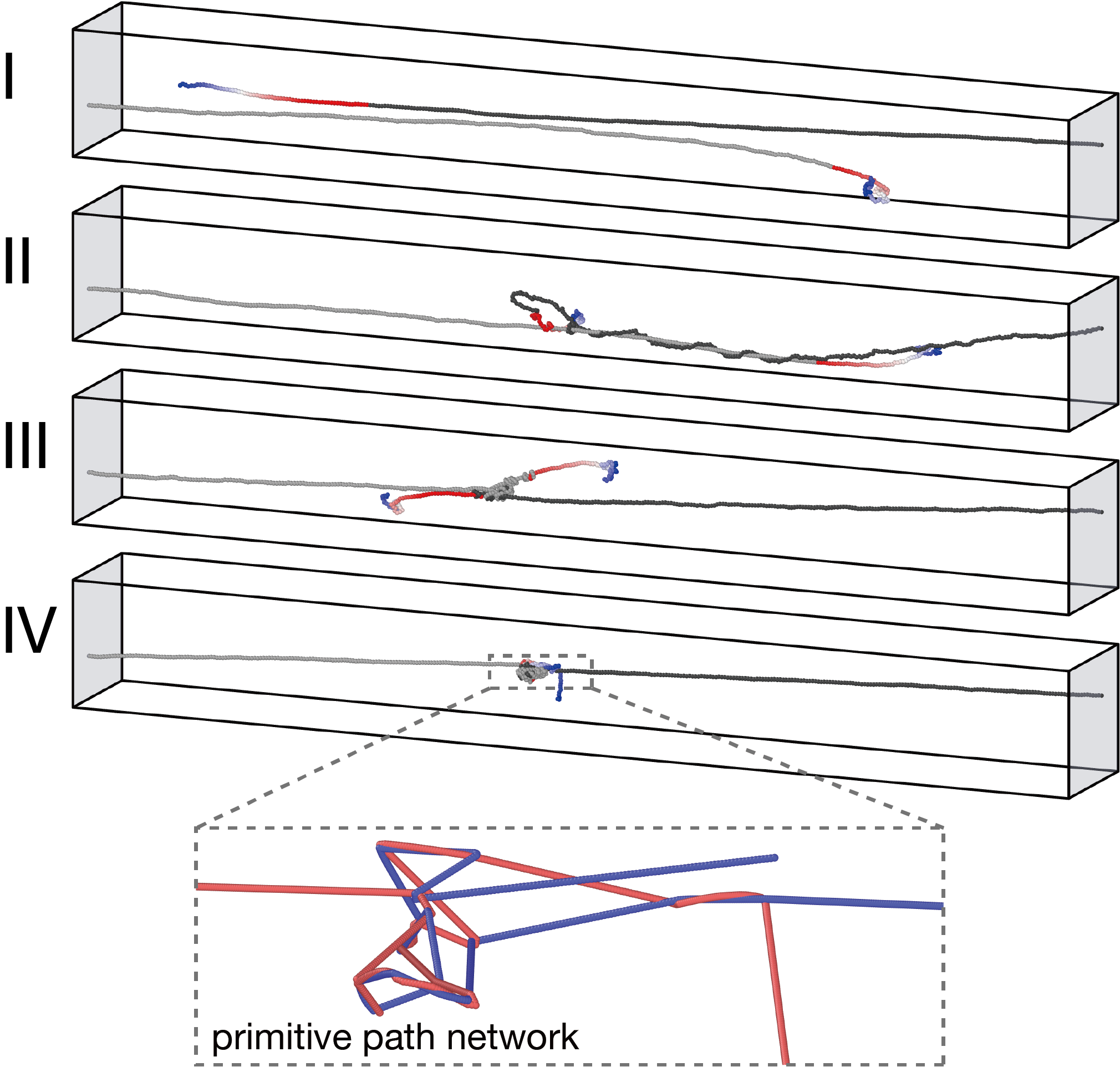}}
		\end{tabular}
	\end{center}
	\caption{\textbf{Inter-molecular bridging knot.} The process of bridging knotting between two anchored passive-active copolymers (from \uppercase\expandafter{\romannumeral1} to \uppercase\expandafter{\romannumeral4}), where the bottom panel is the primitive path network of the knot.}
	\label{fg:load}
\end{figure*}

At last, we consider the interaction between two passive-active copolymer chains anchored separately on opposing surfaces. As shown in Fig.~\ref{fg:load}, when two copolymers are in close proximity, they can quickly twist together and form a compact inter-molecular knot. The probability of forming the knot structure approaches $100\%$ within a relative short time scale ($10^5\tau$) based on $40$ independent simulations. Once the knot structure is formed, it remains stable and compact as long as the active force persists. In order to illustrate the local topology inside the inter-molecular knot, we use CReTA (contour reduction topological analysis) method to extract the primitive path network between the two polymer chains~\cite{Tzoumanekas2006} (see Supplementary Note 4). As shown in the bottom panel of Fig.~\ref{fg:load}, the two chains entangle violently with each other, forming a complex knot with dozens of crossings. Nevertheless, when active forces are withdrawn, this complex knot can unfasten spontaneously. This finding demonstrates the potential applications of actively reptative polymers in macromolecule braiding and weaving~\cite{Qu2021, DabrowskiTumanski2017}.

\section*{Discussion}
In this work, we show that anchored actively reptative polymers with moderate activity can spontaneously form multiple knots by continual self-knotting.  The spontaneous knotting process is caused by the intermittent giant conformation fluctuations and the reptative motion of the chain, which involves the curvature and loop formation of polymer tails. Once a knot is formed, it migrates to the anchoring point by a non-equilibiurm ratchet effect. We have demonstrated that such an active polymer, when graft on the end of a passive macromolecule, can act as an active needle to manipulate the topology of its passive counterpart, by either transferring knots  or directly generating complex knots on the macromolecules. The active needles can also facilitate the bridging knotting between two oppositely anchored passive macromolecules. 

Our work highlights the non-equilibrium effects in modifying the dynamic pathways of polymer chains. Experimentally, the self-knotting phenomenon can be directly verified in colloidal chain systems at the micron scale, where colloidal chain swimmers have already been realized through various methods, such as magnetic driving and catalyst coating~\cite{Dreyfus2005,Biswas2017,Yang2020,Li2016}. At the submicron scale, active chains can be achieved by connecting catalyst/enzyme-modified nanotube motors~\cite{Sanchez2015, Ma2016JACS,Guo2018}. Recently, an enzyme-grafted carbon nanotube with a diameter of about 20 nm was reported to exhibit autonomous propulsion~\cite{Pantarotto2008}, which has the potential to generate knots on chromosome or DNA. In principle, one can also immobilize catalysts/enzymes on a semi-flexible polymer, like double-stranded DNA, to generate self-propelling forces along the backbone~\cite{Ghosh2021,Wilner2009}. Such active molecular chains or needles could be applicable for molecular-level knotting of biomacromolecules. After the formation of the knots, introducing specific bindings between the polymer tail and the anchoring point through controllable methods such as click chemistry is an effective means to permanently preserve the topology of the chain \cite{Haque2020}. We believe the proposed knotting strategy provides a fresh technical routine which opens up many possibilities in macromolecular topology engineering~\cite{Qu2021,DabrowskiTumanski2017}. In addition, our work also inspires the design of innovative microscopic robotic arms (see Fig.~S9 and Movie~S6 for a preliminary demonstration).

\section*{Materials and Methods}
\paragraph{Model.} The active polymer chain we consider consists of $N$  hard-core monomers connected by unbreakable bonds in thermal bath with temperature $T$. One end of the polymer ($i=N$) is anchored on an impenetrable flat surface. Adjacent monomers are connected by finitely extensible nonlinear elastic potential $U_{e}=-150(R_m/\sigma)^2\ln[1-(r/R_m)^2],~r<R_m$, where $R_m=1.05\sigma$ denotes the maximum bond length. $\sigma$ and $r$ are the size of the monomer and the distance between two adjacent monomers, respectively~\cite{Li2023}. The bending energy of the chain is described by a harmonic potential $U_{b}= {K_b}(\pi-\theta)^2/2$ where $K_b$ is the bending stiffness and $\theta$ represents the bond angle. The relative rigidity $\xi$ of the chain is
\begin{eqnarray}
	\xi &=& \frac{L_p}{L}=\frac{2K_b}{Nk_BT},  \label{eq:xi} 
\end{eqnarray}
where $L_p= {2bK_b}/{k_BT}$ is the persistence length with $k_B$ as the Boltzmann constant, and $L=N b$ is the length of the polymer with $b\simeq0.95\sigma$ as the equilibrated bond length. The excluded volume interaction is modelled by a WCA-like potential $U_m=4\varepsilon[(\sigma/r)^{44}-(\sigma/r)^{22}]+\varepsilon,~r<2^{1/22}\sigma$ with $\varepsilon$ being the energy unit of the system. The self-propelling force on monomer $i$ has a constant magnitude along local tangent towards the free end, i.e., $\mathbf{F}_{a,i}=F_{a} {\mathbf{e}_{i-1,i+1}}$ with  $\mathbf{e}_{i,j}=(\mathbf{r}_i-\mathbf{r}_j)/\vert\mathbf{r}_{i,j}\vert$ ($i = 2, 3, \cdots,N-1$).   Here,  $\mathbf{r}_i$ is the  position of monomer $i$. The P\'{e}clet number of the system is defined as
\begin{eqnarray}
	{\rm Pe} &=& \frac{F_a\sigma}{k_BT}. \label{eq:Pe}
\end{eqnarray}
Without losing generality, we set $F_a\sigma=\varepsilon$ and adjust  P\'{e}clet number Pe by changing the temperature $T$. Standard Langevin dynamics is used to simulate the polymers. The velocity Verlet algorithm with a time step of $\Delta t=0.001\tau$ is used to integrate the equations of polymer motion, where $\tau=\sigma/v_0$ is set as the time unit of the system and $v_0=F_a/\gamma$ is the typical velocity of the active reptation. More details about the model and simulations can be found in Supplementary Materials.

\paragraph{Knot topology analysis.} From one end of the polymer, a progressively growing segment is selected and then closed into a ring by adding auxiliary monomers. Subsequently, the closed ring is projected onto a plane to obtain the crossing characteristics, from which the Alexander polynomial can be calculated to determine the knot type. Once a knot is found, reverse scanning from the current location is performed to find the knot with the same Alexander polynomial. Then we get the type, location (starting position of the knot), and size (number of monomers within the knot) of this knot (Fig.~S1). More details about the topology analysis strategy can be found in Supplementary Materials.

\section*{Acknowledgments}
We are grateful to the High-Performance Computing Center of Collaborative Innovation Center of Advanced Microstructures, the High-Performance Computing Center (HPCC) of Nanjing University for the numerical calculations.

\section*{Funding}
This work was supported by the National Natural Science Foundation of China 12104219 (J.X.L.), 12275127 (Q.L.L.), 12174184 (Y.Q.M.), 32301246 (L.L.H.), and 12347102 (Y.Q.M.); the Natural Science Foundation of Jiangsu Province BK20233001 (Y.Q.M.); the Innovation Program for Quantum Science and Technology 2024ZD0300101 (Y.Q.M.).

\section*{Author Contributions}
J.X.L., S.W., Q.L.L. and Y.Q.M. conceived and designed the research. J.X.L. performed the simulations. J.X.L., L.L.H. and Q.L.L. interpreted the results. J.X.L., Q.L.L., and Y.Q.M. wrote the manuscript.

\section*{Competing Interests}
All authors declare that they have no competing interests.

\section*{Data Avilability}
All data needed to evaluate the conclusions in the paper are present in the paper and/or the Supplementary Materials.

\section*{Supplementary Materials}
Supplementary Text; Figs. S1 to S9; Table S1; Movies S1 to S6.\\
Supplementary material for this article is available at https://www.science.org/doi/10.1126/sciadv.adr0716.

\includepdf[pages=-]{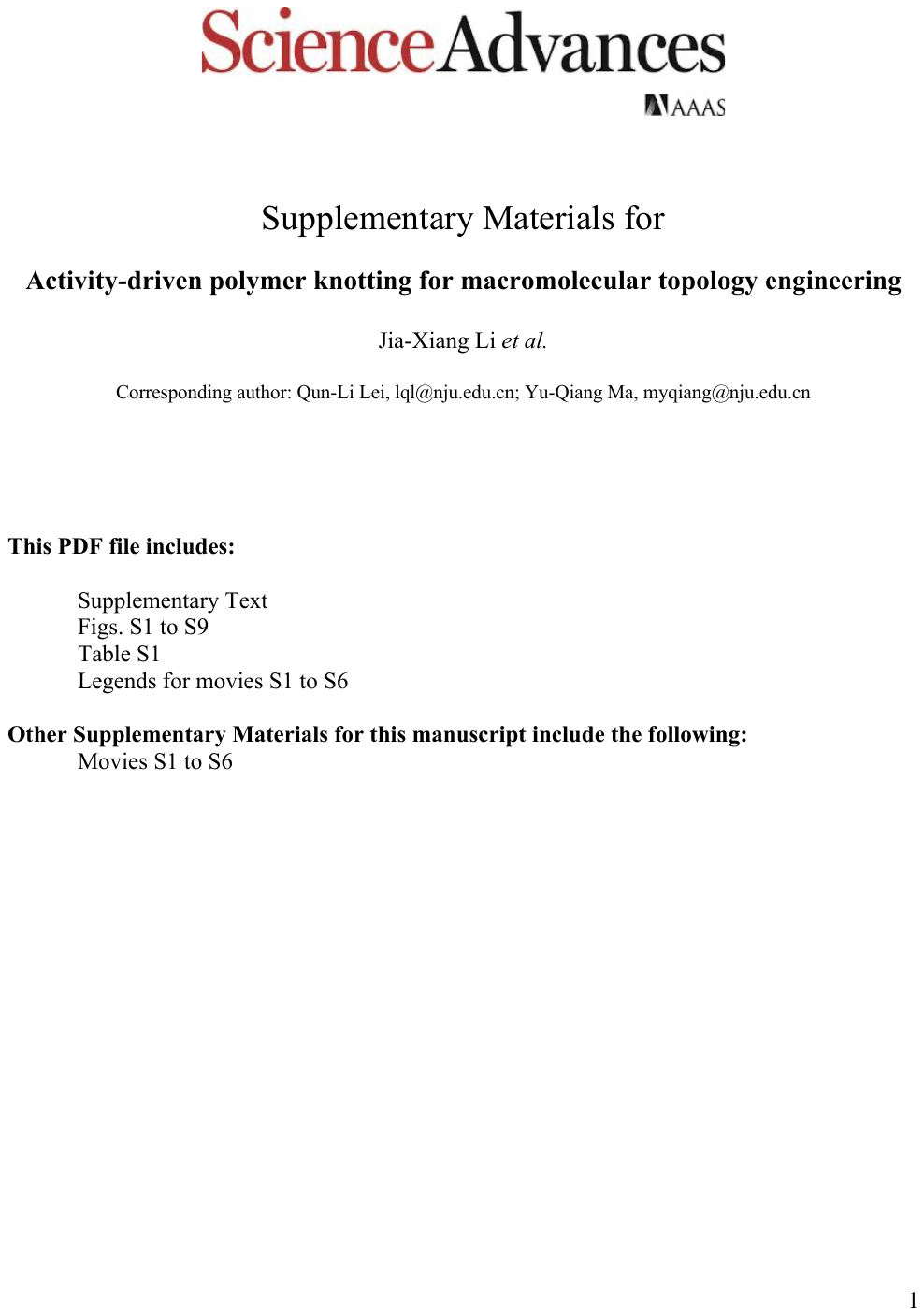}

\end{document}